\documentclass[preprint,prd,showpacs,amsmath,amssymb,amsthm,nofootinbib]{revtex4}
\usepackage[mathscr]{euscript}
\usepackage{bm}
\usepackage{graphicx}
\usepackage{subfigure}
\usepackage{multirow}
\usepackage[colorlinks=true,linkcolor=red]{hyperref}
\newcommand{\bea}{\begin{eqnarray}}
\newcommand{\eea}{\end{eqnarray}}
\newcommand{\beq}{\begin{equation}}
\newcommand{\eeq}{\end{equation}}

\def\/{\over}

\begin{document}

\title{Stability analysis of the Tsallis holographic dark energy model}
\author{Qihong Huang$^{1}$\footnote{Corresponding author: huangqihongzynu@163.com}, He Huang$^2$, Jun Chen$^{3}$, Lu Zhang$^{1}$ and Feiquan Tu$^{1}$}
\affiliation{
$^1$School of Physics and Electronic Science, Zunyi Normal University, Zunyi 563006, China\\
$^2$College of Physics and Electronic Engineering, Nanning Normal university, Nanning 530001, China\\
$^3$School of Science, Kaili University, Kaili, Guizhou 556011, China
}

\begin{abstract}
Using the generalized Tsallis entropy,  the Tsallis holographic dark energy(THDE) was proposed recently. In this paper we analyze the cosmological consequences of the THDE model with  an interaction between dark energy and dark matter $Q=H(\alpha\rho_{m}+\beta\rho_{D})$. We find that the THDE model can explain the current accelerated cosmic  expansion, and it is stable under certain conditions. Furthermore,  through investigating the dynamical analysis,  we find that there exists an attractor which represents an accelerated expansion phase of the universe. When $\beta=0$, this attractor corresponds to a dark energy dominated de Sitter solution and the universe can evolve into an era which is depicted by the $\Lambda$CDM model. The age of universe in this model is also explored.
\end{abstract}

\maketitle

\section{Introduction}

The observational data of Type Ia supernovae~\cite{Perlmutter1999,Riess1998}, the cosmic microwave background radiation~\cite{Spergel2003,Spergel2007} and the large scale structure~\cite{Tegmark2004,Eisenstein2005} have indicated that our universe is undergoing an accelerated expansion. In order to explain this observed scenario, usually, it is assumed that there is an exotic energy component, named dark energy (DE), in our universe. After decades of research, however, the nature of DE is still unclear. A possible candidate of DE is the cosmological constant which is favored by observations~\cite{Planck2015}. However, it suffers from both the fine-tuning problem and the coincidence problem. To explain the accelerating expansion of current universe, other dark energy theories have been proposed, such as, quintessence~\cite{Wetterich1988,Ratra1988,Caldwell1998}, phantom~\cite{Caldwell2002,Caldwell2003}, quintom~\cite{Feng2005,Feng2006,Guo2005}, agegraphic~\cite{Cai2007,Wei2007,Wei2008}, and so on.

An interesting method to explain the origin and nature of DE is applying the holographic principle~\cite{Susskind1995,Bousso2002} into the cosmological framework~\cite{Bak2000,Horava2000}, and then the holographic dark energy(HDE) was proposed~\cite{Cohen1999,Hsu2004,Li2004}. This model is supported by various observations~\cite{Zhang2005,Zhang2007,Huang2004,Enqvist2005,Shen2005}, and  can provide interesting cosmological phenomenology~\cite{Li2004,Huang2004a,Hsu2004,Nojiri2017,Granda2008,Wang2005,Setare2006,Wang2005a,Feng2007,Zimdahl2007,Sheykhi2011,Sheykhi2012}. The cornerstone of the HDE models is the horizon entropy, and any change of the horizon entropy will result in different HDE models. Recently, by considering the Tsallis generalized entropy~\cite{Tsallis2013} and using the Hubble horizon as the IR cutoff, the Tsallis holographic dark energy(THDE) has been proposed. This model can describe the late-time accelerated expansion and predict the age of universe which is in agreement with observations~\cite{Tavayef2018}. Unfortunately, through analyzing the squared sound speed of the THDE model, it is found that the THDE model is unstable at the classical level~\cite{Tavayef2018}. When the THDE was considered in the Brans-Dicke cosmology, the positive squared sound speed can be obtained ~\cite{Ghaffari2018}.

Dynamical system analysis is an excellent method to analyze the qualitative behavior of the nonlinear system. In this method, the field equations are written as an autonomous system, and then the fixed points can be obtained. The stable fixed point is named as the attractor which is used to describe the final state of the universe. The dynamical system analysis has already been used with great success in single-scalar field model~\cite{Copeland1998,Holden1998,Roy2014,Roy2015,Dutta2016,Bhatia2017,Sola2017}, scalar-tensor gravity~\cite{Carloni2008}, f(R) gravity~\cite{Guo2013}, f(T) theory~\cite{Wu2010}, teleparallel dark energy~\cite{Wei2012}, loop quantum gravity~\cite{Fu2008, Wu2008}, mimetic gravity~\cite{Dutta2018}. A general discussion for applying the dynamical system analysis to general relativity and cosmology can be found in~\cite{Bahamonde2018}. For the HDE model, the dynamical system analysis applied to the HDE model with the future event horizon and the Hubble horizon as IR cut off are discussed in~\cite{Setare2009} and~\cite{Banerjee2015}. In~\cite{Banerjee2015}, the phase behavior of the HDE model are analyzed, and the results show that there exists an attractor and the universe will evolve into an accelerated expansion era.

Recently, the stability of THDE model with different IR cutoffs were studied in ~\cite{Zadeh2018}. It was found that the THDE model is unstable at the classical level. However, the situation of $\delta>2$ has not been discussed in the Hubble cutoff, so it remains unclear whether the THDE model is stable and this is exactly what we are planing to address in present paper. Except for studying the stability of the THDE model with the Hubble cutoff under the condition of $\delta>2$, we also analyze its dynamical behavior and the age of the universe dominated by THDE.

The paper is organized as follows. In Section 2, we investigate the evolution of the universe in the THDE model. In Section 3, the phase space behavior of this model is discussed. In Section 4, we analyze the age of the present universe in this model. Finally, our main conclusions are presented in Section 5. Throughout this paper, unless specified, we adopt the metric signature ($-, +, +, +$). Latin indices run from 0 to 3 and the Einstein convention is assumed for repeated indices.

\section{The universe evolution}

By considering the Hubble horizon as the IR cutoff, $L=H^{-1}$, the energy density of the THDE model takes the form of~\cite{Tavayef2018,Tsallis2013}
\bea\label{rD}
\rho_{D}=B H^{-2\delta+4},
\eea
where $B$ is an unknown constant~\cite{Guberina2007,Ghaffari2014,Jahromi2018} and $\delta$ denotes the non-additivity parameter~\cite{Tsallis2013}. The line element for a homogeneous and isotropic flat Friedmann-Robertson-Walker(FRW) universe is given by
\bea
ds^{2}=-dt^{2}+a^{2}(t)[dr^{2}+r^2 d\Omega^{2}],
\eea
where $a(t)$ denotes the scale factor with $t$ being cosmic time. The Friedmann equation has the form
\bea\label{F}
H^{2}=\frac{\kappa^{2}}{3}(\rho_{m}+\rho_{r}+\rho_{D}),
\eea
where, $\kappa^{2}=8\pi G$, $\rho_{m}$, $\rho_{r}$ and $\rho_{D}$ denote the energy density of pressureless matter, radiation and THDE, respectively. The radiation, pressureless matter and THDE conservation equations become
\bea\label{de1}
\dot{\rho_{r}}+4H\rho_{r}=0,
\eea
\bea\label{de2}
\dot{\rho_{m}}+3H\rho_{m}=Q,
\eea
\bea\label{de3}
\dot{\rho_{D}}+3H(1+\omega_{D})\rho_{D}=-Q,
\eea
where $\omega_{D}=\frac{p_{D}}{\rho_{D}}$ is the equation of state parameter of THDE, $Q$ denotes an interaction between THDE and the pressureless matter, while radiation is separately conserved. Throughout this paper, we consider~\cite{Bahamonde2018,Olivares2008,Cabral2009,Quartin2008,Li2010}
\bea
Q=H(\alpha\rho_{m}+\beta\rho_{D}),
\eea
in which $\alpha$ and $\beta$ are coupling constants. When $\alpha=\beta=3b^{2}$ and $\rho_{r}=0$, this model reduces to the case in Ref.~\cite{Zadeh2018}. And the model considered in Ref.~\cite{Tavayef2018} will be recovered for $\alpha=\beta=0$ and $\rho_{r}=0$. Defining the following dimensionless density parameters
\bea
\Omega_{m}=\frac{\kappa^{2}\rho_{m}}{3H^{2}}, \qquad \Omega_{r}=\frac{\kappa^{2}\rho_{r}}{3H^{2}}, \qquad \Omega_{D}=\frac{\kappa^{2}\rho_{D}}{3H^{2}},
\eea
together with the dimensionless variable
\bea
\sigma=\frac{\kappa^{2}Q}{3H^{3}}=\alpha \Omega_{m}+\beta \Omega_{D},
\eea
we find that the Friedmann equation~(\ref{F}) can be rewritten as
\bea\label{T1}
\Omega_{m}+\Omega_{r}+\Omega_{D}=1.
\eea

Differentiating the Friedmann equation~(\ref{F}) and using Eqs.~(\ref{de1}),~(\ref{de2}),~(\ref{de3}),~(\ref{T1}), one can obtain
\bea\label{HH}
\frac{\dot{H}}{H^{2}}=\frac{1}{2}\Big[\Omega_{m}+(1-3\omega_{D})\Omega_{D}\Big]-2.
\eea
Taking the time derivative of Eq.~(\ref{rD}) and using Eqs.~(\ref{de3}) and (\ref{HH}), we have
\bea\label{oD}
\omega_{D}=\frac{[(\delta-2)(\Omega_{m}+\Omega_{D})+5-4\delta]\Omega_{D}-\sigma}{3[(\delta-2)\Omega_{D}+1]\Omega_{D}}.
\eea
For $\delta<1$, we obtain $\delta-2<-1$ which means that there exists a divergence in the evolution of $\omega_{D}$ when $\Omega_{D}=\frac{1}{2-\delta}$. Thus, $\delta<1$ will lead to a singularity in the evolution of $\omega_{D}$~\cite{Tavayef2018}. Throughout the paper, we consider the situation $\delta\geq 1$.

Combining above Eqs.~(\ref{HH}) and ~(\ref{oD}), one can calculate the deceleration parameter $q$
\bea
q=-1-\frac{\dot{H}}{H^{2}}=\frac{2-(1+\alpha)\Omega_{m}-(\beta+2\delta)\Omega_{D}}{2[1+(\delta-2)\Omega_{D}]}.
\eea

Defining $\Omega'=d\Omega/d(lna)$ and using Eqs.~(\ref{de2}),~(\ref{de3}),~(\ref{T1}),~(\ref{HH}), we obtain
\bea\label{Om}
\Omega^{'}_{m}=[(3\omega_{D}-1)\Omega_{D}-\Omega_{m}+1]\Omega_{m}+\sigma,
\eea
\bea\label{OD}
\Omega^{'}_{D}=[(3\omega_{D}-1)(\Omega_{D}-1)-\Omega_{m}]\Omega_{D}-\sigma.
\eea

\begin{figure}[!htb]
                \centering
                \includegraphics[width=0.457\textwidth ]{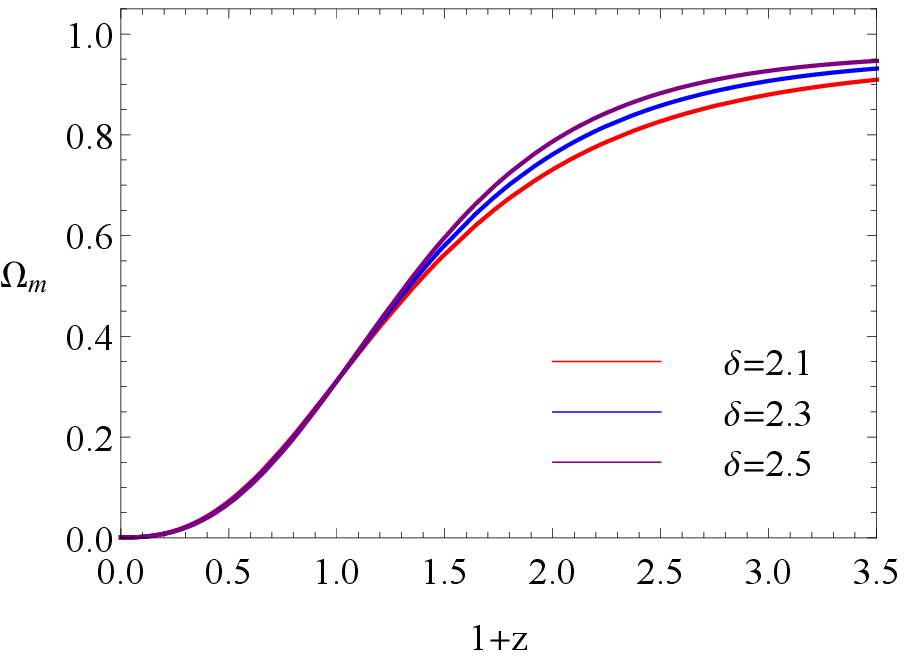}
                \includegraphics[width=0.457\textwidth ]{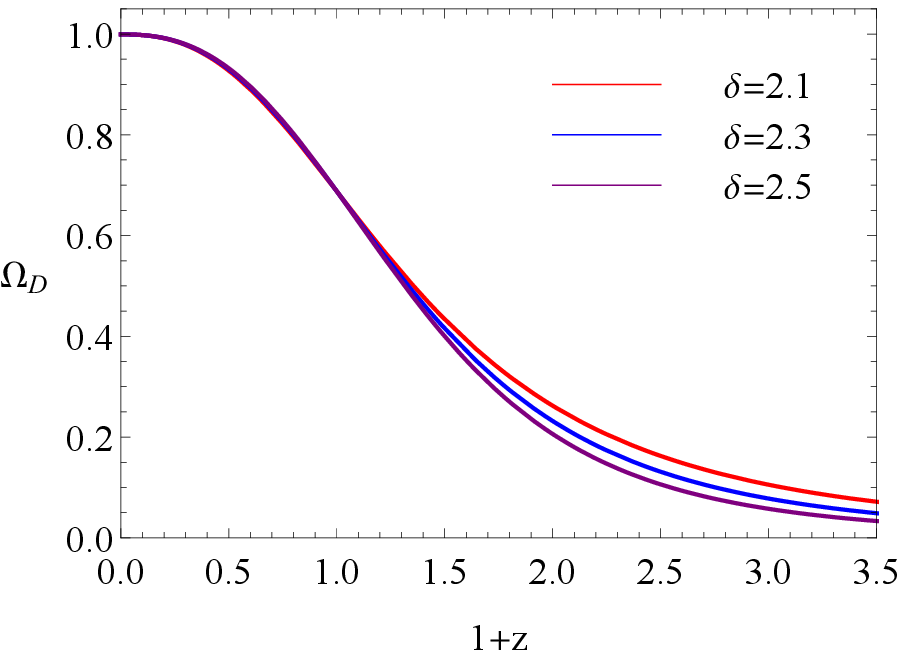}
                \caption{\label{Fig1} The evolution of $\Omega_{m}$ and $\Omega_{D}$ versus redshift parameter $z$ with $\alpha=0.5$, $\beta=0.002$, $\Omega^{0}_{m}=0.311$ and $\Omega^{0}_{D}=0.688$.}
\end{figure}

\begin{figure}[!htb]
                \centering
                \includegraphics[width=0.457\textwidth ]{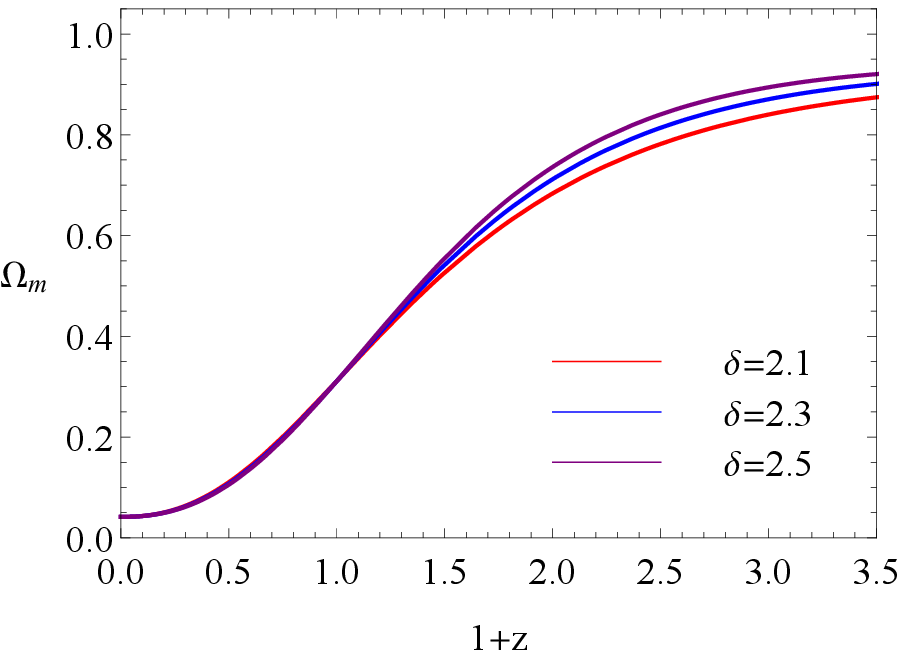}
                \includegraphics[width=0.457\textwidth ]{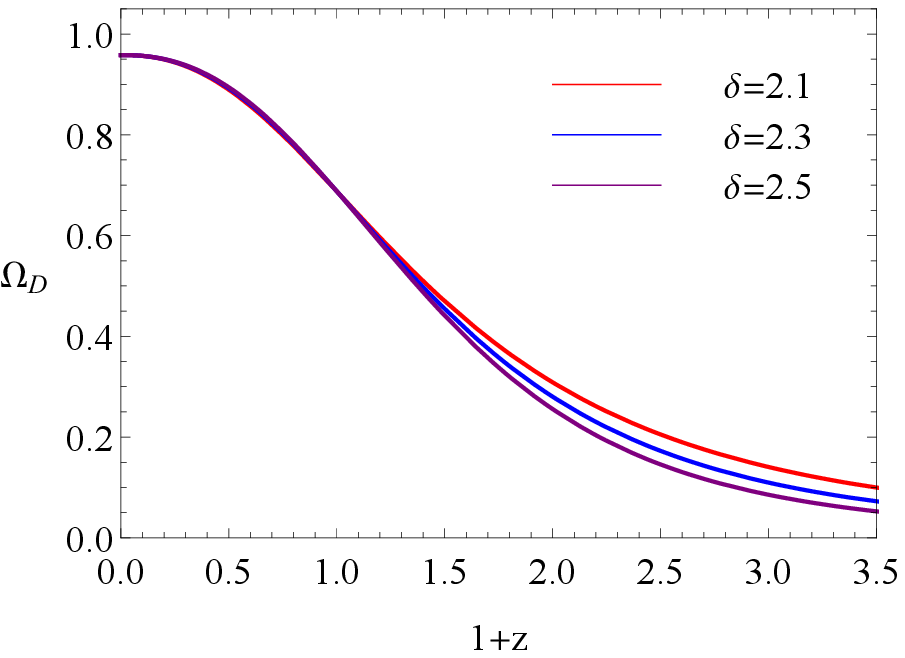}
                \caption{\label{Fig2} The evolution of $\Omega_{m}$ and $\Omega_{D}$ versus redshift parameter $z$ with $\alpha=0.7$, $\beta=0.1$, $\Omega^{0}_{m}=0.311$ and $\Omega^{0}_{D}=0.688$.}
\end{figure}

\begin{figure}[!htb]
                \centering
                \includegraphics[width=0.457\textwidth ]{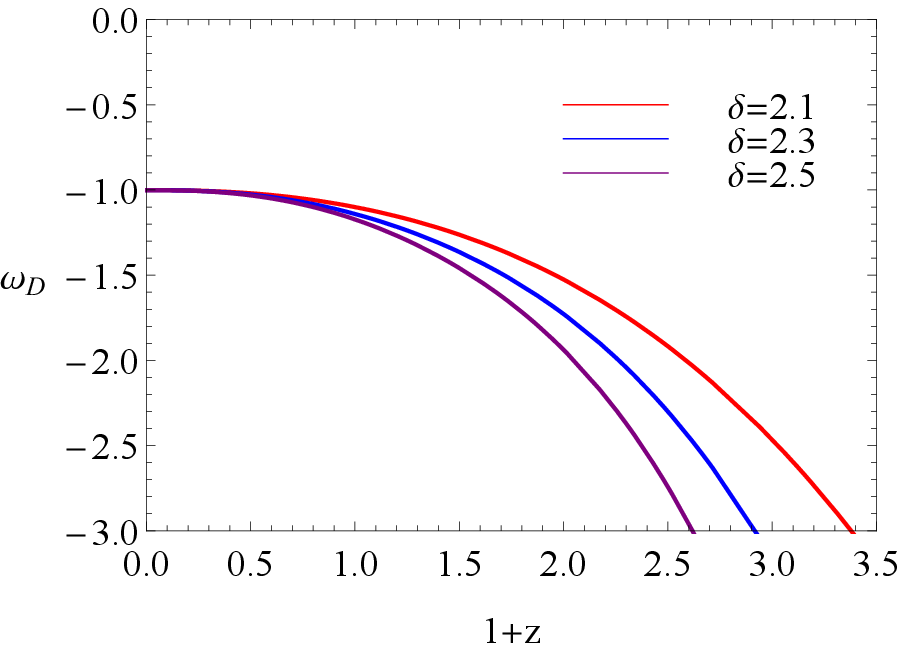}
                \includegraphics[width=0.457\textwidth ]{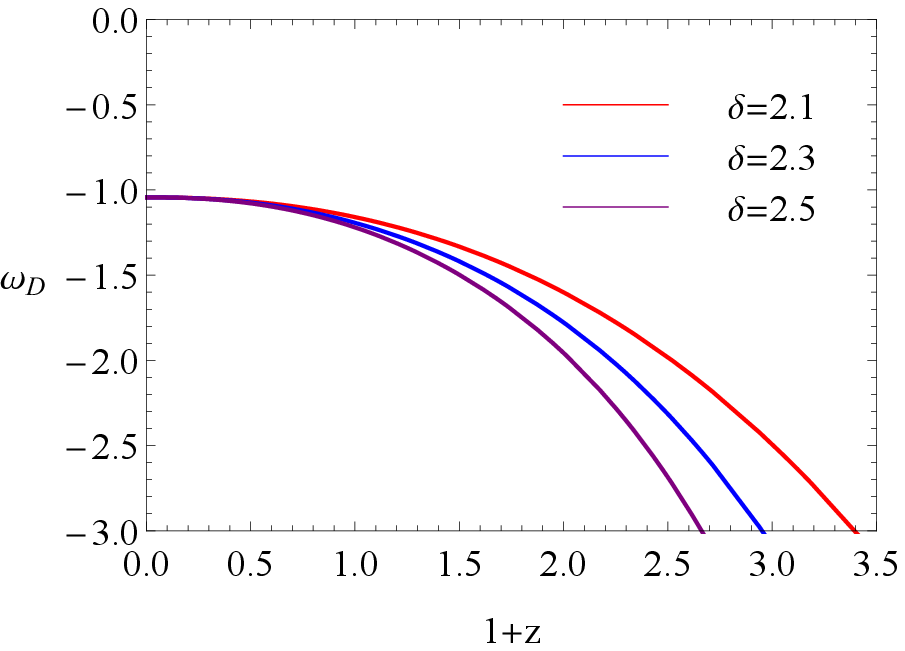}
                \caption{\label{Fig3} The evolution of $\omega_{D}$ versus redshift parameter $z$ with $\Omega^{0}_{m}=0.311$ and $\Omega^{0}_{D}=0.688$. The left panel is plotted for $\alpha=0.5$, $\beta=0.002$, while the right one is plotted for $\alpha=0.7$, $\beta=0.1$.}
\end{figure}

\begin{figure}[!htb]
                \centering
                \includegraphics[width=0.457\textwidth ]{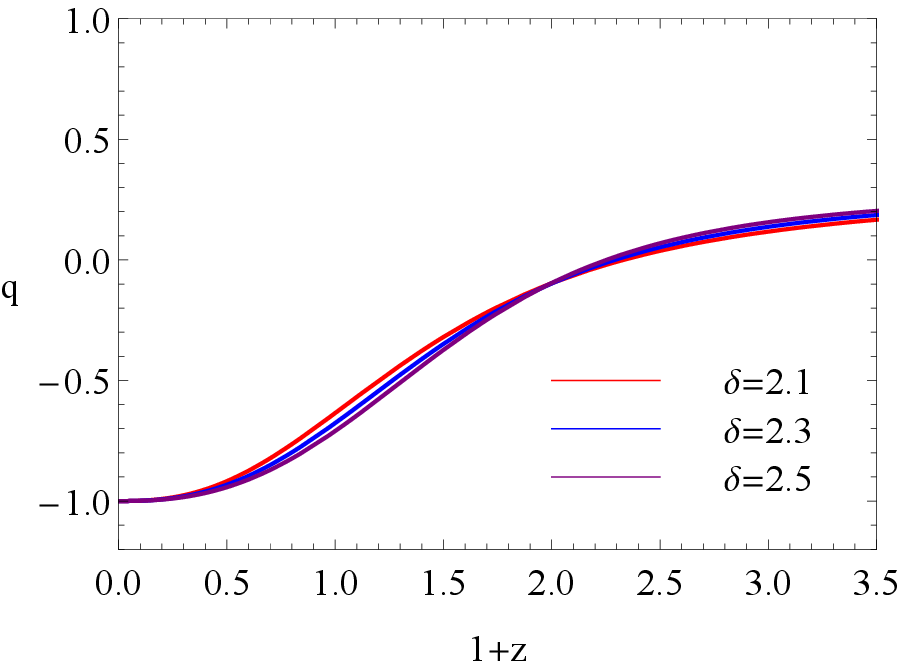}
                \includegraphics[width=0.457\textwidth ]{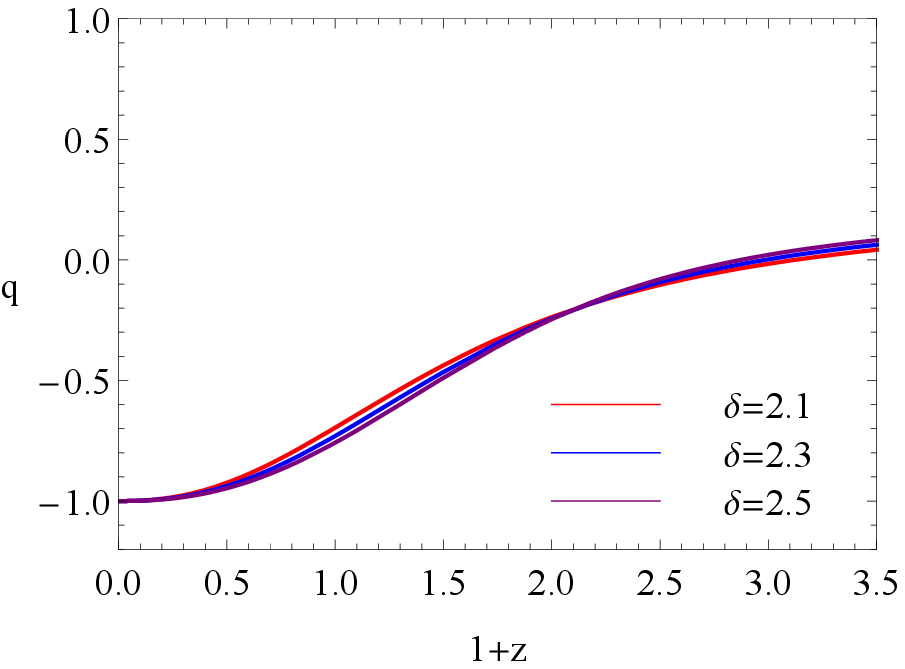}
                \caption{\label{Fig4} The evolution of $q$ versus redshift parameter $z$ with $\Omega^{0}_{m}=0.311$ and $\Omega^{0}_{D}=0.688$. The left panel is plotted for $\alpha=0.5$, $\beta=0.002$, while the right one is plotted for $\alpha=0.7$, $\beta=0.1$.}
\end{figure}

For $\alpha=0.5$, $\beta=0.002$ and $\alpha=0.7$, $\beta=0.1$ with the initial condition $\Omega^{0}_{m}=\Omega_{m}(z=0)=0.311$, $\Omega^{0}_{D}=\Omega_{D}(z=0)=0.688$~\cite{Planck2015}, the evolution of $\Omega_{m}$, $\Omega_{D}$, $\omega_{D}$ and $q$ versus $(1+z)$ are shown in Figs.~\ref{Fig1}, ~\ref{Fig2}, ~\ref{Fig3} and ~\ref{Fig4}. Fig.~\ref{Fig1} and ~\ref{Fig2} indicate that in the early time($z\rightarrow\infty$) we have $\Omega_{D}\rightarrow 0$, while at the late time($z\rightarrow -1$) we have $\Omega_{m}\rightarrow 0$ and $\Omega_{D}\rightarrow 1$. From Fig.~\ref{Fig3}, one can see that the model behaves as the phantom source, and $\omega_{D}=-1$ is obtained when $\beta=0$ at the late time. The deceleration parameter $q$ has been plotted for some values of $\delta$, $\alpha$ and $\beta$ in Fig.~\ref{Fig4}. It is obvious that a desired asymptotic behavior($q=-1$) is obtained at the late time, and our universe undergoes a transition from the deceleration phase to an accelerated one during the evolution.

\begin{figure}[!htb]
                \centering
                \includegraphics[width=0.457\textwidth ]{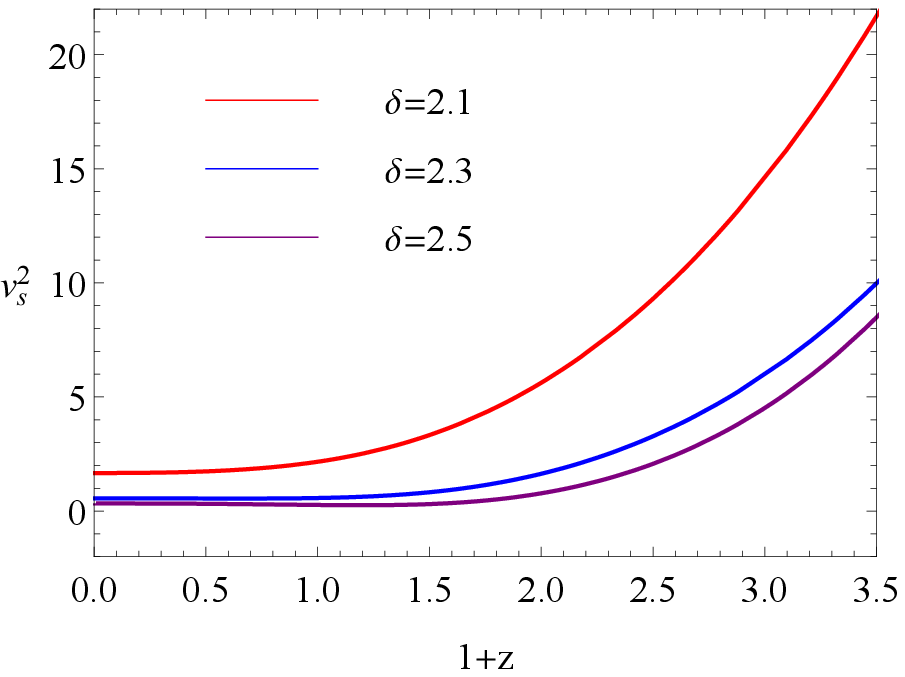}
                \includegraphics[width=0.457\textwidth ]{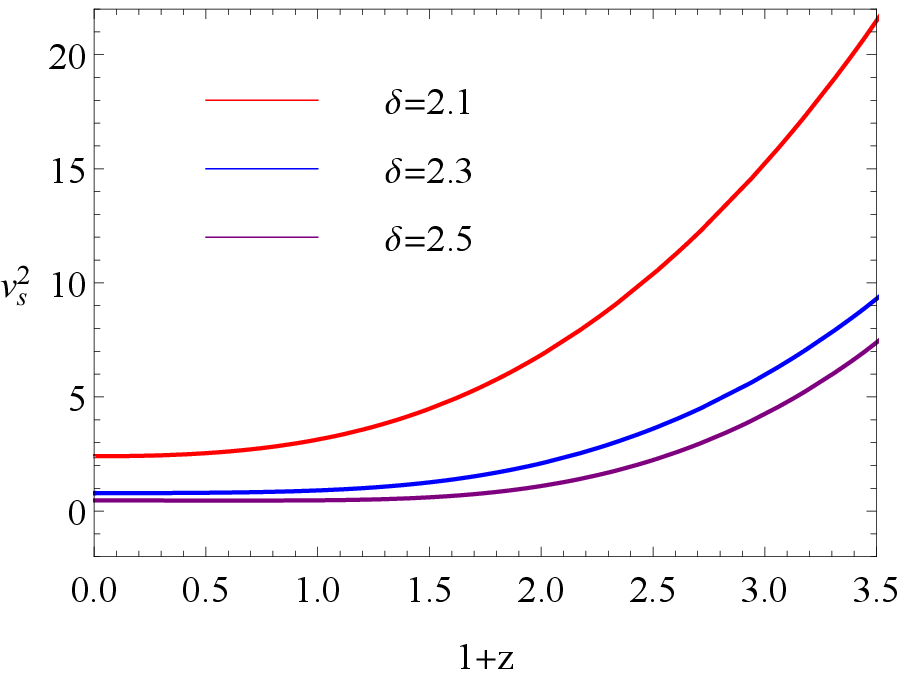}
                \caption{\label{Fig5} The evolution of $v^{2}_{s}$ versus redshift parameter $z$ with $\Omega^{0}_{m}=0.311$ and $\Omega^{0}_{D}=0.688$. The left panel is plotted for $\alpha=0.5$, $\beta=0.002$, while the right one is plotted for $\alpha=0.7$, $\beta=0.1$.}
\end{figure}

To explore the stability of the THDE model against perturbation, we will analyze its squared sound speed $v^{2}_{s}$. For $v^{2}_{s}>0$, the model is stable against perturbation. Otherwise, it is unstable. The squared sound speed is defined as
\bea
v^{2}_{s}=\frac{dp_{D}}{d\rho_{D}}=\frac{\dot{p}_{D}}{\dot{\rho_{D}}}=\frac{\rho_{D}}{\dot{\rho_{D}}}\dot{\omega}_D+\omega_{D},
\eea
which leads to
\bea\label{vs2}
v^{2}_{s}=\frac{1}{3}\Bigg[\frac{4(\Omega_{m}+\Omega_{D}-1)+\frac{\alpha[(\alpha-3)\Omega_{m}+\beta \Omega_{D}]}{(\delta-2)\Omega_{D}}}{(\alpha+1)\Omega_{m}+(4+\beta)\Omega_{D}-4}+\frac{(\alpha+1)(\delta-1)\Omega_{m}+(\Omega_{D}-1)[4(\delta-1)+\beta]}{[(\delta-2)\Omega_{D}+1]^{2}}\Bigg],
\eea
where Eqs.~(\ref{rD}),~(\ref{HH}),~(\ref{oD}),~(\ref{Om}) and~(\ref{OD}) are employed. To discuss the stability of this model, we will study the value of $v^{2}_{s}$. Since it is difficult to obtain the general conditions for $v^{2}_{s}>0$, we will consider a special situation, i.e. both the first and second term of Eq.~(\ref{vs2}) are positive. Then, we can obtain the conditions for $v^{2}_{s}>0$. We do not show all stable conditions here because they are too complicated, and find that one of the stable conditions is
\bea
&&2<\delta\leq5,\ 0<\Omega_{m}<1,\ 0<\Omega_{D}<1-\Omega_{m},\ \alpha>\frac{4(1-\Omega_{D})-\Omega_{m}}{\Omega_{m}},\nonumber\\
&&0\leq\beta<\frac{[(\alpha+1)\Omega_{m}-4(1-\Omega_{D})](\delta-1)}{1-\Omega_{D}}.
\eea
when considering $\alpha\geq0$ and $\beta\geq0$. It is obvious that this model can be stable under perturbations, which can also be found in Fig.~\ref{Fig5} where the evolutions of the squared sound speed $v^{2}_{s}$ with some values of $\delta$, $\alpha$ and $\beta$ are shown.

For the case of $\alpha=\beta=3b^{2}$ and $\rho_{r}=0$, $v^{2}_{s}$ can reduce to the result in Ref.~\cite{Zadeh2018}
\bea\label{vs2b}
v^{2}_{s}=\frac{(\delta-1)(\Omega_{D}-1)+b^{2}\big[\delta+\frac{1}{(\delta-2)\Omega_{D}}\big]}{[1+(\delta-2)\Omega_{D}]^{2}},
\eea
and the positive squared sound speed requires
\bea
\delta>2,\ 0<\Omega_{D}\leq 1,\ b>\sqrt{\frac{(\delta-2)(\delta-1)(1-\Omega_{D})\Omega_{D}}{1+\delta(\delta-2)\Omega_{D}}}.
\eea
Thus, one can see that the model is stable for $\delta>2$, but this case is not analyzed in Ref.~\cite{Zadeh2018}. However, for this case $\delta>2$, $\omega_{D}$ will approach to $-1$ when $b^{2}$ approaches $0$, but $\omega_{D}$ can not equal to $-1$ since $b^{2}>0$ in this model. While in our model with $\beta=0$, $\omega=-1$ can be realized in the late time, and the universe can evolve into the dark energy dominated epoch which is depicted by the $\Lambda$CDM model.

\section{Phase space analysis of the model}

In previous section, we have discussed the evolution of the universe with the initial conditions $\Omega^{0}_{m}=0.311$ and $\Omega^{0}_{D}=0.688$ in THDE model and found that our universe undergoes a transition from the deceleration phase to an accelerated one during the evolution. In order to discuss the dynamical evolution, now we analyze the phase space behavior of the system. To achieve this goal, we consider it in an autonomous system with Eqs.~(\ref{Om}) and~(\ref{OD}). Following e.g.~\cite{Wang2012,Banerjee2015}, the critical points of the autonomous system can be obtained by taking
\bea
\Omega^{'}_{m}=\Omega^{'}_{D}=0.
\eea
We obtain three critical points ($P_{1}$, $P_{2}$ and $P_{3}$) and their existence conditions which are given in Table~\ref{Tab1}. The values of $\Omega_{r}$ for points $P_{1}$, $P_{2}$ and $P_{3}$ are 1, 0, 0, respectively. Thus, the first critical point $P_{1}(0,0)$ corresponds to the early radiation dominated era, the second point $P_{2}(1,0)$ is matter dominated, and point $P_{3}(\frac{\beta}{3-\alpha+\beta},\frac{3-\alpha}{3-\alpha+\beta})$ represents an accelerated expansion phase of the universe since the deceleration parameter is $q=-1$. For point $P_{1}$ and $P_{2}$, they are always exist, but the corresponding equation of state parameter $\omega_{D}$ is divergent. For point $P_{3}$, its coordinate and the equation of state parameter $\omega_{D}=\frac{3-\alpha+\beta}{\alpha-3}$ are full determined by the value of $\alpha$ and $\beta$. If $\beta=0$, one obtains $\omega_{D}=-1$ in the late time and point $P_{3}$ becomes dark energy dominated de Sitter solution.

\begin{table}
\caption{\label{Tab1} The critical points and their existence conditions of the autonomous system.}
\begin{center}
 \begin{tabular}{|c|c|c|c|}
  \hline
  \hline
  $Label$ & $Critical \ Points(\Omega_{m}, \Omega_{D})$ & $Existence$ & $q$\\
  \hline
  $P_{1}$ & $(0,0)$ & $Always$ & $1$\\
  \hline
  $P_{2}$ & $(1,0)$ & $Always$ & $\frac{1-\alpha}{2}$\\
  \hline
  $P_{3}$ & $(\frac{\beta}{3-\alpha+\beta},\frac{3-\alpha}{3-\alpha+\beta})$ & $0\leq\alpha<3,\beta\geq0 \ or \ \alpha>3,\beta=0$ & $-1$\\
  \hline
  \hline
  \end{tabular}
\end{center}
\end{table}

Now, we analyze the stabilities of these critical points. After linearizing the autonomous system, one gets two differential equations. The stability of these critical points is determined by the eigenvalues of the coefficient matrix of the differential equations. If the eigenvalue of the coefficient matrix is negative, the corresponding critical point is stable. Otherwise, it is unstable. The existence conditions are considered when we analyze the stabilities of these critical points. Eigenvalues and their stability conditions of these critical points are given in Table~\ref{Tab2}. Here, the condition $\delta\neq2$ is used because Eq.~(\ref{vs2}) indicates that $v^{2}_{s}$ is divergent for $\delta=2$. The results in Table~\ref{Tab2} show that points $P_{2}$ and $P_{3}$ can be stable under some special conditions, but they can not be stable simultaneously. However, for points $P_{1}$ and $P_{2}$, the equation of state parameter $\omega_{D}$ is divergent, and these points are not suitable. So, the stable critical point is point $P_{3}$, which requires $0\leq\alpha<3$ and $1<\delta<2$ or $\delta>2$. Under these conditions, only point $P_{3}$ is stable. Thus, there exists only one attractor, which is an accelerated expansion solution, for this autonomous system.

\begin{table}
\caption{\label{Tab2} The eigenvalues of critical points and their stability conditions.}
\begin{center}
 \begin{tabular}{|c|c|c|c|}
  \hline
  \hline
  $Label$ & $Eigenvalues$ & $Conditions$ & $Points$\\
  \hline
  \multirow{1}*{${P_{1}}$}
  &\multirow{1}*{$1+\alpha, 4(\delta-1)$} & $\alpha\geq0, 1<\delta<2 \ or \ \delta>2$ & $Unstable \ point$\\
  \hline
  \multirow{2}*{${P_{2}}$}
  &\multirow{2}*{$-(1+\alpha), -(\delta-1)(\alpha-3)$} & $\alpha>3, 1<\delta<2 \ or \ \delta>2$ & $Stable \ point$\\
  \cline{3-4}
  & & $0\leq\alpha<3, 1<\delta<2 \ or \ \delta>2$ & $Saddle \ point$\\
  \hline
  \multirow{2}*{${P_{3}}$}
  &\multirow{2}*{$-4, \frac{(\alpha-3)(\delta-1)(\alpha-3-\beta)}{(\alpha-3)(\delta-1)-\beta}$} & $0\leq\alpha<3, \beta\geq0, 1<\delta<2 \ or \ \delta>2$ & $Stable \ point$\\
  \cline{3-4}
  & & $\alpha>3, \beta=0, 1<\delta<2 \ or \ \delta>2$ & $Saddle \ point$\\
  \hline
  \hline
  \end{tabular}
\end{center}
\end{table}

\begin{figure}[!htb]
                \centering
                \includegraphics[width=0.457\textwidth ]{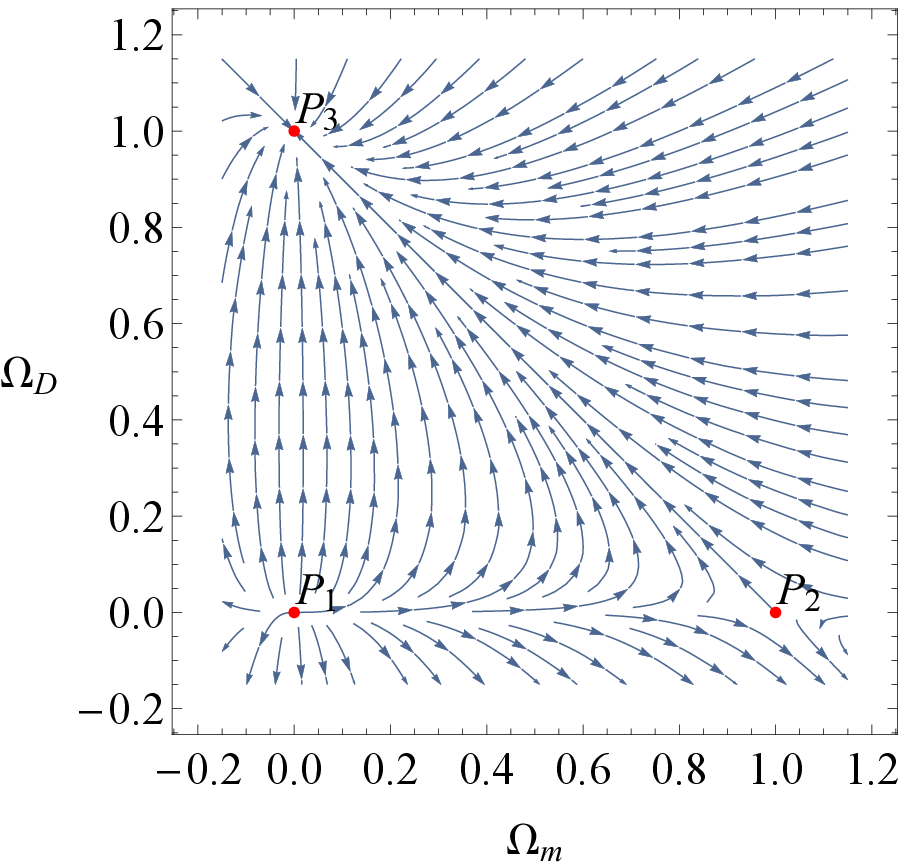}
                \includegraphics[width=0.412\textwidth ]{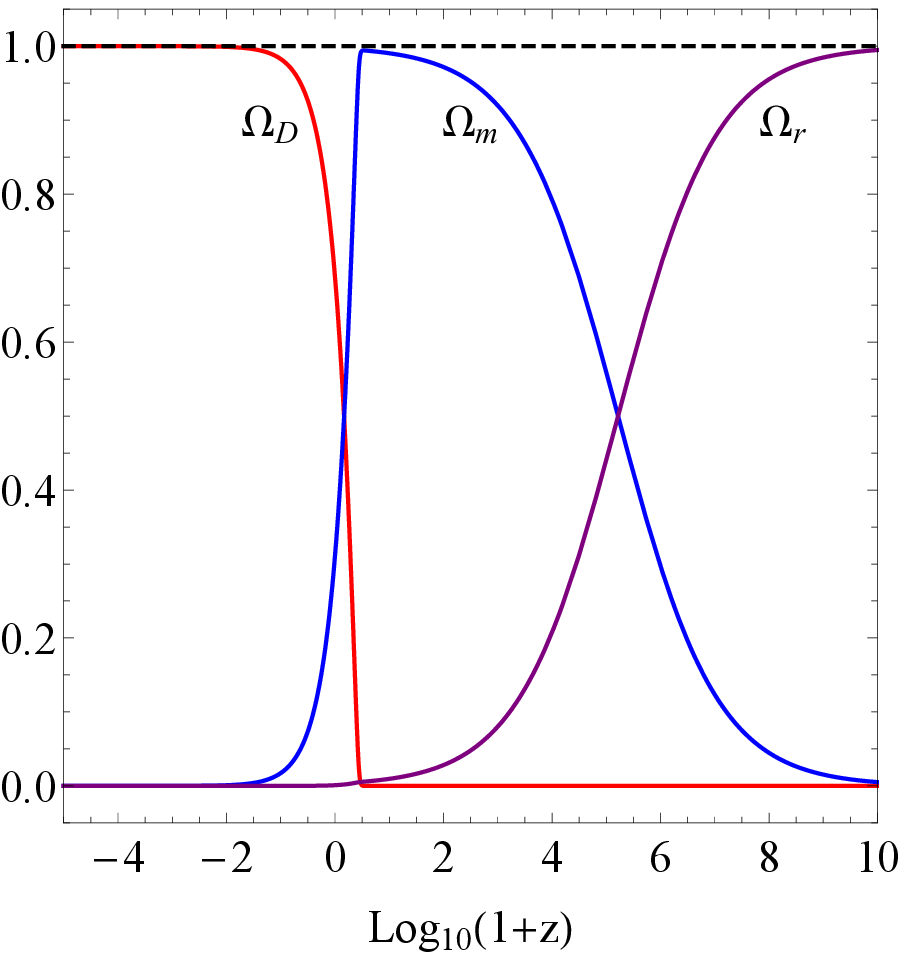}
                \caption{\label{Fig6} The left panel is the phase diagram of ($\Omega_{m},\Omega_{D}$) with $\delta=2.1$, $\alpha=0.5$ and $\beta=0$. The right panel is the evolution of $\Omega_{r}$, $\Omega_{m}$ and $\Omega_{D}$ which are plotted for $\delta=25$, $\alpha=0.1$ and $\beta=0$.}
\end{figure}

For $0\leq\alpha<3$ and $1<\delta<2$ or $\delta>2$, $P_{2}$ is the saddle point and $P_{3}$ is the stable point. In the left panel of Fig.~\ref{Fig6}, we plot a phase diagram of ($\Omega_{m},\Omega_{D}$) for this situation. This figure shows that all curves in this phase diagram will converge into the stable point $P_{3}$ which is an attractor, and our universe will finally evolve into an accelerated expansion phase. The right panel of Fig.~\ref{Fig6} shows the evolutions of $\Omega_{r}$, $\Omega_{m}$ and $\Omega_{D}$. These figures depict the expansion history of the universe, i.e. the universe starts from the early radiation dominated phase (the unstable point $P_{1}$), passes through the matter dominated epoch (the saddle point $P_{2}$) and finally evolves into a late-time accelerated expansion epoch (the stable point $P_{3}$).

For $\beta=0$, the stable point $P_{3}$ becomes $(0,1)$ with $\omega=-1$ which represents a de Sitter expansion solution. In this case, our universe will eventually evolve into a dark energy dominated epoch which is depicted by the $\Lambda$CDM model.

\section{The universe age}

It is well known that our universe cannot be younger than its constituents~\cite{Alcaniz1999}. If a model which predicts the universe's age is less than its constituents, this model suffers from the so-called age problem. In order to discuss the age problem, we estimate the order of the age of the current universe($z=0$). The age of the present universe can be written as
\bea
t=\int \frac{dt dH}{dH}=\int \frac{1}{\frac{\dot{H}}{H^{2}}}\frac{dH}{H^{2}}\approx\bigg(\frac{1}{\frac{\dot{H}}{H^{2}}}\bigg)_{z=0}\int \frac{dH}{H^{2}}=\frac{2}{4-(\Omega^{0}_{m}+(1-3\omega_{D})\Omega^{0}_{D})}\frac{1}{H_{0}},
\eea
where $H_{0}$ is the present value of the Hubble parameter. If we consider $\delta=2.1$, $\alpha=0.5$ and $\beta=0.002$, we have $t=\frac{2.74}{H_{0}}$. In another case with $\delta=2.1$, $\alpha=0.7$ and $\beta=0.1$, we have $t=\frac{3.28}{H_{0}}$. For the oldest star $HD 140283$, its age was estimated to be $14.46\pm 0.8$ Gyr~\cite{Bond2013}. And the Planck 2015 results~\cite{Planck2015} predict the age of the universe is $13.799$ Gyr. It is obvious that this THDE model does not suffer from the age problem.

\section{Conclusion}

Based on the generalized Tsallis entropy, the THDE model was proposed. Although it can describe the late-time accelerated universe, it is unstable against perturbations. In this paper, by considering the Hubble cutoff and the interaction $Q=H(\alpha\rho_{m}+\beta\rho_{D})$ between THDE and pressureless matter, we analyze the evolution of the THDE and investigate their cosmological consequences. We find that the case $\delta>2$ can produce suitable behavior for $\Omega_{D}, \omega_{D}$, and $q$. The stability analysis of the squared sound speed shows that the THDE model is stable against perturbations under certain conditions. Through the dynamical analysis of the THDE model, we find that the system has a stable fixed point which is an attractor and represents an accelerated expansion phase of the universe since $q=-1$. This attractor represents the dark energy dominated epoch. When $\beta=0$, it corresponds to a de Sitter expansion solution and the universe will eventually evolve into an era which is depicted by the $\Lambda$CDM model. Finally, we investigate the age of the universe in this model and find that this model does not suffer from the age problem.

\begin{acknowledgments}

We would like to thank P. Wu for his help. This work was supported by the National Natural Science Foundation of China under Grants Nos. 11865018, 11865019, 11847031, 11847085, 11465011, the Foundation of the Guizhou Provincial Education Department of China under Grants Nos. KY[2018]312, KY[2018]028, KY[2017]247, KY[2016]104, the Doctoral Foundation of Zunyi Normal University of China under Grants No. BS[2017]07.

\end{acknowledgments}

\end{document}